# A CITY-CENTRIC APPROACH TO ESTIMATE AND EVALUATE GLOBAL URBAN AIR MOBILITY DEMAND

Lukas Asmer[1] (https://orcid.org/0000-0002-5975-5630),
Roman Jaksche[1] (https://orcid.org/0000-0001-9607-9283),
Henry Pak[1] (https://orcid.org/0000-0002-6259-3441)
Petra Kokus[1] (https://orcid.org/0000-0002-2132-9572)

[1]DLR Institute of Air Transport
German Aerospace Center
Linder Höhe, 51147 Cologne, Germany

Contact: lukas.asmer@dlr.de

**Abstract**

Urban Air Mobility (UAM) is expected to effectively complement the existing transportation system by providing fast and safe travel options, contributing to decarbonization, and providing benefits to citizens and communities. A preliminary estimate of the potential global demand for UAM, the associated aircraft movements, and the required vehicles is essential for the UAM industry for their long-term planning, but also of interest to other stakeholders such as governments and transportation planners to develop appropriate strategies and actions to implement UAM. This paper proposes a city-centric forecasting methodology that provides preliminary estimates of the potential global UAM demand for intra-city air taxi services for 990 cities worldwide. By summing all city-specific results, an estimate of the global UAM demand is obtained. By varying the parameters of the UAM system, sensitivity studies and different market scenarios are developed and analyzed. Sensitivity analyses show how strongly demand decreases when air taxi ticket prices increase. Considering low ticket prices and high vertiport densities, possible market development scenarios show that there is a market potential for UAM in over 200 cities worldwide by 2050. The study highlights the significant impact of low ticket prices and the need for high vertiport densities to drive UAM demand. This highlights the need for careful optimization of system components to minimize costs and increase the quality of UAM services.

**Keywords**
Urban Air Mobility, Global UAM demand, Market potential, Market development

**NOMENCLATURE**

| | |
|---|---|
| AMT | Alternate Mode of Transportation |
| EASA | European Aviation Safety Agency |
| EU | European Union |
| eVTOL | electric Vertical Take Off and Landing |
| GC | Generalized Cost |
| GDP | Gross Domestic Product |
| Km | Kilometer |
| KPI | Key Performance Indicator |
| NASA | National Aeronautics and Space Administration |
| OD | Origin-Destination |
| OEM | Original Equipment Manufacturer |
| PKM | Passenger Kilometers |
| SF | Scaling Factor |
| SLF | Seat Load Factor |
| Sq. km | Square kilometer |
| UAM | Urban Air Mobility |
| USA | United States of America |
| U.S. | United States |
| VD | Vertiport Density |
| VTT | Value of Travel Time |

## 1. INTRODUCTION

Due to the increasing advances in new vehicle concepts and technologies, Urban Air Mobility (UAM) is expected to effectively complement the existing transportation system by providing fast and safe travel options, contributing to decarbonization, and providing benefits to citizens and communities. The concept of UAM is not entirely new. A first wave of urban air mobility took place between the 1950s and the 1980s, mainly enabled by the emergence of turbine-powered helicopters. But the use of helicopters as an urban transportation mode failed to take off due to a lack of profitability and social acceptance [1-3]. Currently, however, the integration of UAM into existing urban transportation systems as a complementary component is becoming more and more conceivable.

The term UAM covers many several applications to meet different transport needs, such as intra-city, airport shuttle or suburban commuter [4]. In general, the term UAM is associated with an air transportation system based on a high-density vertiport network and air taxis services within an urban environment, enabled by new technologies and integrated into multimodal transportation systems. The transportation is performed by electric aircraft taking off and landing vertically, remotely piloted or with a pilot on board [5]. UAM has the potential to offer various advantages and benefits for different stakeholders. In this respect, UAM is expected to enable safer, cleaner and faster mobility within urban agglomerations. Studies have shown that the use of air taxis can save 15 to 40 minutes on an average standard urban travel time [5]. The European Aviation Safety Agency (EASA) considers UAM as a new safe, secure and more sustainable air transportation system for passengers and cargo in urban environments, enabled by new technologies and integrated into multimodal transportation systems. However, the introduction of UAM is associated with technological, regulatory, infrastructural, social and economic challenges, which require a holistic approach as well as close cooperation between the various stakeholders in order to unlock the full potential of this new mode of transportation. In this process, the UAM system components must be designed to ensure that the system is accessible, affordable, safe, secure, and sustainable for users as well as profitable for operators [6].

While initial eVTOL manufacturers plan to have UAM vehicle certification by 2023 and start first operations in 2024 [7-9], the global market potential of UAM is still unclear. A preliminary estimate of the potential UAM demand, the associated number of flight movements, and the required number of vehicles would be helpful for e.g. manufacturers to plan upcoming production in advance. At the same time, the estimates can be useful for authorities, service providers or research institutions to assess the impact and effects of a potential UAM development from an overall system perspective at an early stage.

One of the main challenges in estimating the potential demand for UAM is that cities and urban agglomerations differ in various aspects (area, population, geographic characteristics, wealth level, cultural background, etc.). From a global perspective, the development of specific transport models for each city, taking into account all transport-related parameters (e.g. mobility patterns, existing transportation infrastructure or transportation policies), is not feasible due to time and cost constraints.

This paper proposes a city-centric forecasting methodology based on a limited set of input parameters relevant for UAM to provide first estimates of the potential global UAM demand, aircraft movements and fleet size for intra-city air taxi services. This model is part of a holistic view of UAM, and is among a set of forecasting models that exist for each of the above use cases.

The remainder of the paper is structured as follows:

Section 2 provides a review of the research project's background. The importance of forecasting methods for estimating the global UAM demand is explored, along with an overview of existing literature in the field. Section 3 outlines the methodology and underlying assumptions of the study, to ensure transparency and replicability. The steps taken to estimate the global UAM demand for intra-city air taxi services are described in detail. The results of the research are presented in section 4. 990 cities worldwide were analyzed using the methodology. The section contains sensitivity analysis as well as multiple market development scenarios highlighted as part of the study. Section 5 contains the conclusion. In this section, the main research findings are summarized, the limitations of the approach are highlighted, and potential future research is discussed. This section emphasizes the significance of the present work and its potential impact on further UAM development.

## 2. BACKGROUND

Demand forecasting is an essential component in designing the efficiency, sustainability, and profitability of new transportation systems. In particular, when evaluating new modes of transportation, the ability to accurately predict future demand enables various stakeholders to make informed decisions, optimize operations, reduce environmental impacts, and increase customer satisfaction. Forecasting capability is also critical for the evaluation and impact assessment of UAM as a novel urban transportation system. A preliminary estimate of the potential demand for UAM, the associated number of aircraft movements, and the number of vehicles required is fundamental to the further design of UAM. This will help stakeholders to develop appropriate strategies and actions to maximize the benefits of UAM while addressing potential challenges. Thus, a global UAM forecast can help plan vehicle production according to demand, use resources efficiently, coordinate transportation effectively, integrate ground infrastructure according to demand in cities, better support environmental goals, and set appropriate frameworks and safety standards.

However, forecasting the global UAM demand involves a number of challenges. As long as there are no UAM systems, it is not possible to base estimates of future development on historical data. Uncertainties also remain in connection with technological development and market launch, which makes it difficult to make reliable long-term forecasts.

Urban transportation systems are extremely complex, multi-faceted and individual, as cities have different characteristics such as population size, built-up area or wealth level, which have a direct impact on the people's mobility behavior in each city and ultimately on the potential demand for UAM. To address all these heterogeneous market conditions in one approach, a method is needed that is as simple and transferable as possible for all cities without the need to create individual, city-specific transport models.

Currently, many international research groups are working on different aspects of UAM in order to find optimal solutions for the implementation of UAM. However, regarding the preliminary estimation of global UAM demand, there is currently little research available. Initial market studies were carried out several years ago by consulting companies such as Roland Berger [10], Horvath and Partner [11], Porsche Consulting [12] and KPMG [13]. These studies primarily provide an overview of the potential opportunities and economic benefits that could result from the introduction of UAM. However, they tend to provide less insight into the underlying methodologies and assumptions, making it difficult to transfer the methodologies and results.

On the other hand, there are a number of publications in the scientific literature that present concrete methodological approaches to determine UAM demand and potential more precisely. These papers usually provide detailed insights into the models, assumptions and data base used. They enable in-depth analysis and a better understanding of the factors influencing UAM demand. In order to address the complexity of heterogeneous market conditions, these UAM forecasts use an approach that groups cities into clusters and conducts detailed analyses for a representative city in each cluster.

Mayaconda et al. (2020) [14] provide a top-down methodology to estimate the UAM demand which is applied to 31 cities around the world. Based on travelers' willingness to pay for UAM service, the potential UAM traffic volume is estimated. The studies were conducted for the reference year 2035.

Anand et al. (2021) [15] provide a scenario-based evaluation of global UAM demand. The research is based on the previous approach of Mayaconda et al. (2020) and is applied to 542 cities worldwide to determine the global UAM demand. In addition, a scenario-based forecasting approach is used to provide long-term market demand for low and high penetration of UAM services for a 2035-2050 timeframe.

Straubinger et al. (2021) [16] also propose a scenario-based estimation of the global UAM demand. The analysis covers the four dimensions: UAM use cases, city archetypes, market development scenarios, and time horizons from 2020 to 2050, and takes into account different market penetration rates for UAM that vary by the four dimensions, covering uncertainties in prices, travel speeds, network density, access times, and mode choice behavior. By applying the market penetration rates to the conditions of the considered cities, a specific demand is calculated and compared with results from other studies.

Furthermore, there are several studies that have been conducted for smaller geographic areas.

Particularly well-known is the study by Booz Alleen Hamilton [17] commissioned by NASA. They examined the air taxi demand for 10 U.S. cities at a very detailed level, taking into account not only potential demand but also possible systemic constraints such as willingness to pay, infrastructure capacity, time of day, and weather constraints.

A second study commissioned by NASA [18] investigated the market potential of last-mile delivery, air metro services and air taxi services for different U.S. cities. Taking into account the target markets, consumers' willingness to pay and the availability of the technology, demand was determined. By multiplying the total number of expected trips in each city by the percentage of trips eligible for UAM, the market size was calculated.

Rihmja et al. (2022) [19] conducted a demand estimation and feasibility assessment for UAM in Northern California Area. A sensitivity analysis was conducted to examine the impact of cost per passenger mile and number of vertiports on UAM demand. The spatial distribution of UAM demand in the region is also analyzed, with the San Francisco Financial District identified as a major attraction for commuter trips. The results show that low UAM fares and comparable reliability with car travel are necessary to achieve sufficient demand for commuter trips and reduce empty flights.

EASA [20] identified suitable UAM cities in Europe which are the most attractive EU urban target markets for UAM OEMs and UAM operators. The study was conducted for the different sub-use-cases of Urban Air Mobility: airport shuttle, sightseeing, fixed metropolitan network, first aid, medical supply delivery and last-mile delivery. The ranking is based on key performance indicators (KPIs), an infrastructure feasibility assessment for the considered use cases, and a timing feasibility assessment.

Ploetner et al. (2020) [21] investigated the market potential of UAM for public transport in the Munich metropolitan area. An existing agent-based transport model was extended by socio-demographic changes until 2030 and the integration of intermodal UAM services. To simulate the demand for UAM, an incremental logit model was developed. The study defines three UAM networks with different numbers of vertiports and performs sensitivity analyses on factors such as fare, vehicle speed, passenger check-in times at transfer stations, and network size.

Pertz et al. (2022) [22] developed an approach for modeling the UAM commuter demand in Hamburg, Germany. The approach is based on a discrete choice model that predicts commuters' mode choice. For this purpose, predefined traffic cells are used to generate and distribute a door-to-door commuter traffic. By combining the modal split and the market volume for commuting, the market share for commuter UAM traffic in Hamburg is determined. The model offers the possibility to evaluate individual passenger routes and catchment areas as well as to analyze characteristics of travelers and routes with high and low demand.

## 3. METHODOLOGY

This paper proposes a forecasting methodology to provide initial estimates of the potential global UAM demand for intra-city air taxi services following the ideas of the traditional four-step transportation model. The four-step model [23] is a widely used approach for the determination of total and mode-specific transport demand, among others. It requires a detailed database including information on population, household size and income, activity patterns, on existing and future transportation infrastructure, supply, pricing etc. At a global stage, these data are not available at a sufficient level of detail. Therefore, a city-centric approach (**FIG 1**) was developed that uses a limited number of parameters to estimate the UAM demand for a city or an urban agglomeration. The characteristics of the city that serve as the main input parameters of the city-centric approach are the number of inhabitants, the urban area, and the country-specific GDP per capita as a proxy for the level of wealth. These data are available for the present and for the future, some of them publicly and some of them commercially. Demographia's World Urban Areas Database [24] serves as database that covers 990 urban agglomerations with more than 500,000 inhabitants and provides the number of inhabitants, the population density and the built-up urban area of each agglomeration in the year 2022. GDP per capita (real, harmonized) is taken from [25]. The number of inhabitants and GDP per capita in the future is determined by applying country-specific growth rates of population [25] and of GDP per capita by [25].

By applying this method to a set of worldwide cities and summing up all city-specific results, an estimate of global UAM demand for intra-city air taxi services is provided. Variation of major characteristics of the UAM transportation system allows different scenarios to be developed and analyzed.

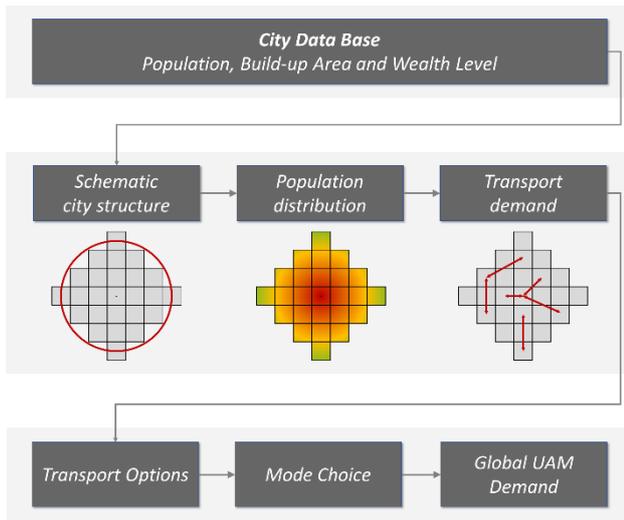

**FIG 1** Concept of the city-centric forecasting approach.

UAM transport demand is derived in five main steps:

**Schematic city structure**

First, each city structure is mapped into a circular structure consisting of square grid cells. Using Hamburg in Germany as an example, **FIG 2** shows how the generic circular city shape is generated from the real extension of the city by transforming it into a grid cell structure. The actual shape of the city is omitted and instead approximated to a circle using the city area and a predefined grid cell size which is the same for all cities. Thus, the individual number of grid cells depends on the city area and the grid cell size.

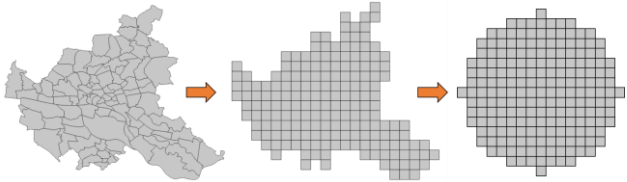

**FIG 2** From original city shape to generic shape – Example of the City of Hamburg, Germany

**Population distribution**

Second, the population is distributed among the grid cells with the highest population density in the center grid cell. The population density decreases from the center grid cell with increasing distance to the city outskirts. The distribution of the population is based on two universal patterns which can be observed worldwide: the greater the distance from the city center, the lower the density, and the larger the city, the more distance is needed for the population density to decrease [26]. The population distribution is achieved by using equation (1). It determines a population density factor $p$ for each grid cell based on its distance $d$ from the center grid cell in relation to the maximum distance $d_{max}$ between the center and the outer grid cell. The factor $x$ indicates the ratio between the population density in the city center and in the outer grid cell, while the value $k$ is a reference for the population density in the outer grid cell of the city and is used to shape the progression of the equation.

$$p(d) = e^{(\ln(x*k) - \ln(k)) * \frac{dmax - d}{d} + \ln(k)} \quad (1)$$

To determine the population density for each grid cell, the total population of the city is distributed among the grid cells in proportion to the size of the individual density factor $p$.

For all cities examined, it is assumed that the population decreases by a factor $x=10$ and the reference value $k=2$, from the center to the edges of the city (**FIG 3**). Very similar developments can be observed in different cities around the world. In North American, East Asian, and Pacific cities with populations larger than 10 million, the size assumption of this factor is best expressed, while it varies with population size and world region [26].

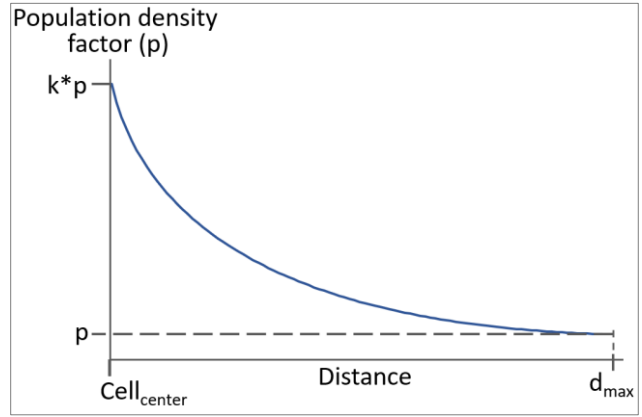

**FIG 3** Population density factor versus distance from the city center

**Transport demand**

Third, a trip table (OD matrix) is constructed, indicating the number of trips between each pair of grid cells. For this purpose, an average daily trip rate per person is used to determine the number of trips originating from a cell. It is assumed that each person makes on average three trips per day, with no distinction made by trip purpose, household income, sex or age. The assumption is based on the "Mobility in Germany" report [27], although this figure may change in relation to socio-demographic characteristics in different countries and cities [28, 29].

Then the resulting trips are distributed to all other cells by using an empirical trip length distribution which is based on GPS car movement data from the U.S. metropolitan region of Dallas, TX [30], shown in **FIG 4**. Only trips that start and end within the city limits are considered in the examination. Trips that go beyond the city limits are out of scope. The diagram reveals that 99 percent of the total population's trips are made up to a distance of 100 km. This entails the characteristic that trips in cities with smaller areas are not completely covered within the city boundaries, but additionally go beyond them. For example, in a city with a maximum distance of 40 km within the urban area, almost 82% of trips are made inside the city boundaries and 18% of trips go beyond them.

Since there are only discrete distances due to the grid cell structure, a special procedure is developed to distribute the trips. Based on the GPS data, equation (2) is determined, which calculates the proportion of trips from the cities' total number of trips as a function of the discrete distances $x_i$.

$$y(x_i) = 0.2051 * \log(x_i) + 0.0592 \quad (2)$$

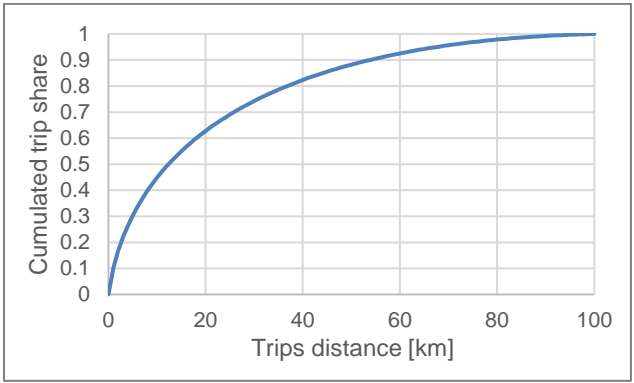

**FIG 4** Empirical trip length distribution of the U.S. metropolitan region of Dallas, TX

Due to the symmetric shape of the generic city and the arrangement of grid cells, single discrete distances occur between multiple pairs of grid cells. Therefore, the number of total trips per discrete distance must be divided among the corresponding pairs of grid cells. Once this procedure is completed, a predefined percentage of outbound trips is determined for each grid cell. Trips remain within a grid cell for a discrete distance equal to half the length of a grid cell edge. This predefined percentage of outbound trips is then adapted according to the population of the destination grid cell. For equal discrete distances between two or more pairs of cells with the same origin cell, the population size in the destination cell affects how high the trip number is on each pair of cells. In this way, the attractiveness of grid cells is highlighted and considered by grid cell specific characteristics, which in this case is expressed by the size of the population. This allocation represents the final step of the trip distribution calculation that leads to the final OD matrix.

**Transport options**

The fourth step is to create the transport options. It is assumed that in addition to the air taxi, there is an alternative mode of ground transportation (AMT) that represents the mode that is currently being used (**FIG 5**). For each OD pair, travel times and travel cost are determined for both modes of transportation.

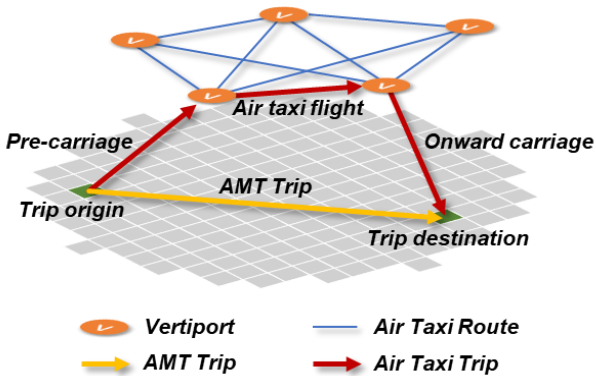

**FIG 5** Itineraries for air taxi and for alternative mode of transportation

The alternate mode trip consists of a direct connection between origin and destination. The associated travel time is calculated based on the linear distance between origin and destination and assuming a constant average speed of 18 km per hour [31]. The monetary cost of an alternate mode trip is calculated by using a price per km and a detour of 20 percent. The price per km varies from country to country depending on many factors, such as market prices for the vehicle, maintenance and insurance, vehicle age, energy consumption or the structure of charges and taxes [32]. Costs are determined based on results of the EU-funded project COMPETE (**FIG 6**), which analyzed the average operating costs per pkm by car in the EU and the USA taking into account key macroeconomic indicators such as information on national fleet structure, average fuel consumption, GDP per capita adjusted for purchasing power, national interest rates and different degrees of liberalization [32]. As the analysis was already carried out in 2014, the operating costs were adjusted to current market conditions [33].

The costs per km of the alternate mode are calculated by:

$$Costs_{amt} = \left(6*10^{-6} * (GDP\ per\ capita_{country}) + 0.0703\right) * 1.7 \qquad (3)$$

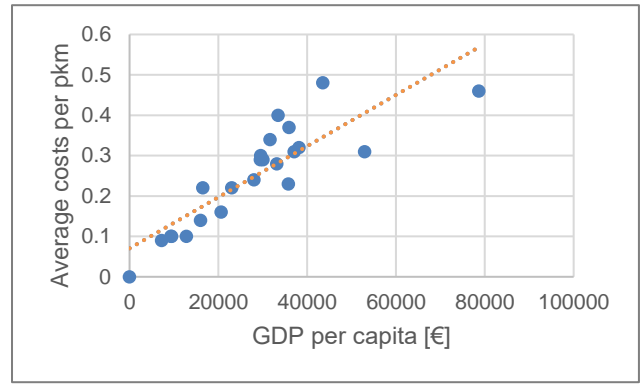

**FIG 6** Average costs per pkm of the alternate mode versus GDP per capita

The air taxi trip consists of three segments: pre-carriage, air taxi flight, and onward carriage. In order to model air taxi trips, first vertiports are evenly distributed by placing them using the sunflower algorithm (**FIG 7**) [34].

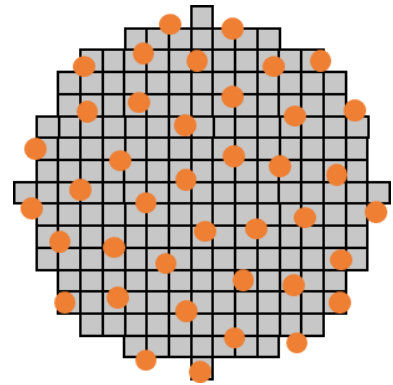

**FIG 7** Schematic city with 253 grid cells and 40 vertiports

The number of vertiports to be placed depends on the city area and a prescribed vertiport density $vd_{city}$. Vertiport density varies from city to city, dependent on the city area and GDP per capita, and is determined for each city individually. It is plausible that cities with lower levels of wealth have difficulty building high-quality transportation systems, resulting in lower vertiport density. Additionally, smaller cities typically have shorter distances to be covered, thus lower vertiport density is expected.

For this purpose, a vertiport density $vd_{ref}$ is assumed, which can be understood as a target value and is valid for cities with an area larger than 3000 sq. km and a GDP per capita larger than that of the United States. Cities where the area and GDP per capita are greater or equal to the refence values are assigned the vertiport density of the reference city. For all other cities, the vertiport density is scaled down by using a scaling factor for the area ($SF_{area}$, **FIG 8**) and for the GDP per capita ($SF_{GDP}$, **FIG 9**), considering the different city characteristics. The scaling functions are designed in a way that a small deviation in GDP per capita has a significantly larger impact on the scaling factor than the area of the city, which only has an influence when the city has only around 1/5 of the area of the reference city. Depending on how the urban area and GDP per capita of the city under consideration differ from the values for the reference city, the reference vertiport density is adjusted resulting in a city-specific vertiport density $vd_{city}$:

$$vd_{city} = vd_{ref} \cdot SF_{area} \cdot SF_{GDP} \tag{4}$$

where $SF_{area}$ and $SF_{GDP}$ are the scaling factors, and $vd_{ref}$ is the reference vertiport density.

Furthermore, it is required that the number of vertiports of a city is at least five. The reference value of the vertiport density can be varied to investigate the impact on the air taxi demand.

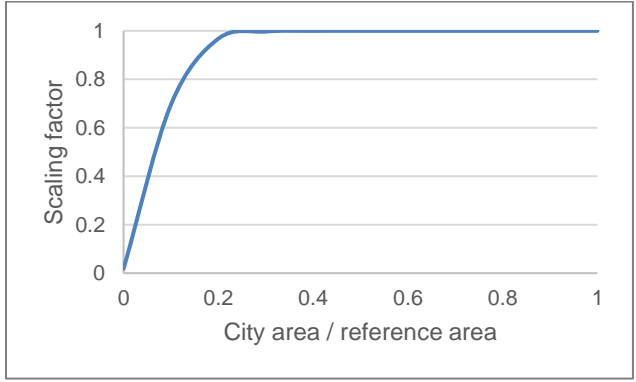

**FIG 8** Function for determining the scaling factors for the area

The pre-carriage from the origin to the next vertiport and the onward carriage from the arrival vertiport to the destination are performed by the alternate mode. The monetary cost and time of travel by air taxi are composed of the cost and time of the flight and of the ground transportation to and from the vertiport. The cost and time for the flight are calculated based on the linear distance between the departure and the arrival vertiport and considering a detour of 5 percent. Cruise speed is set to 100 km/h. Time for take-off and landing is taken into account and set to 2 minutes each. In addition, 3 minutes each are added for boarding and deboarding. Travel costs for the flight are determined using an air taxi ticket price per km, which is constant for all countries and can be varied to study its effect on air taxi demand. The cost and time for the first and last mile are determined based on the linear distance between origin and departure vertiport resp. between arrival vertiport and destination, using the ticket price per km, the speed, and the detour factor of the alternate mode.

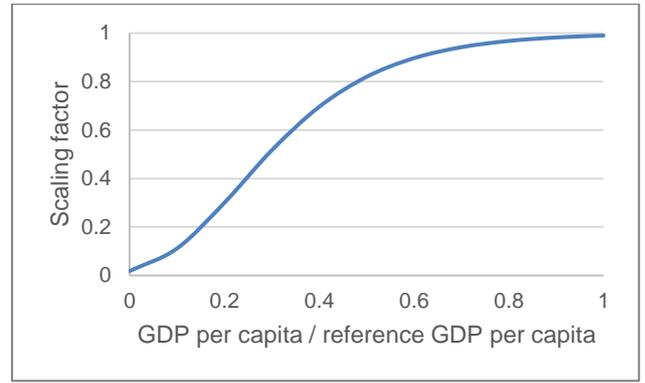

**FIG 9** Function for determining the scaling factors for the GDP

**Mode Choice**

In the last step, a simplified multinomial logit model is applied to determine the mode choice between the alternate mode of transportation and the air taxi, considering the total transport demand and the characteristics of transport options between each pair of grid cells. The air taxi share of each OD pair $i$-$j$ is calculated by ($i$ and $j$ omitted for simplicity):

$$P_{air\ taxi} = \frac{e^{U_{air\ taxi}}}{e^{U_{air\ taxi}} + e^{U_{amt}}} \tag{5}$$

where $U_{air\ taxi}$ and $U_{amt}$ are the utilities of the air taxi and the alternate mode. The utilities are calculated as:

$$U_m = \beta_m + \beta_{GC} \cdot GC_m \tag{6}$$

where $U_m$ is the utility of mode $m$, $\beta_m$ is the mode-specific constant, and $\beta_{GC}$ is the parameter associated with the generalized cost of travel $GC_m$ of mode $m$.

The mode-specific constant can be interpreted as the average effect of unincluded factors on the utility of an option relative to the other alternatives [35]. Since there is currently no reliable information on whether users tend to prefer the air taxi over the car or not, it is first assumed that $\beta_{amt} = \beta_{air\ taxi} = 0$. This assumes that mode choice depends only on travel time and cost. $\beta_{GC}$ generally varies depending on factors such as trip purpose and car availability [36]. In the context of this work, which refrains from further differentiation of trips and travelers, we obtained reasonable results with values of $\beta_{GC}$ between -0.2 and -0.3.

Generalized costs involve variable monetary costs of a trip and monetized travel time:

$$GC_m = c_m + VTT \cdot t_m \tag{7}$$

where $GC_m$ is the generalized cost of travel from origin $i$ to destination $j$ with mode $m$, $c_m$ and $t_m$ are the costs and the time associated with the trip, and $VTT$ is the value of travel time.

$VTT$ is the monetary value of travel time reduction, in other words: how much it is worth to a user to reduce travel time by a time unit, usually by an hour. Since travel time can be thought of as a disutility, consumers are generally willing to pay to have less of it. $VTT$ can vary significantly depending on many different factors, such as the purpose of the trip, the traveler's income, the mode of transportation or the weather conditions, and is usually different for each person [37]. On a global scale, it is challenging to distinguish $VTT$ taking into account these specific factors due to limited data availability. Therefore, $VTT_{city}$ is calculated as a function of the GDP per capita:

$$VTT_{city} = 0.0003 * (GDP\ per\ capita_{country}) - 0.3404 \quad (8)$$

Equation (8) is based on an extensive meta-analysis by Wardman et al. (2016) [38] that considered 3109 monetary valuations from 389 European studies conducted between 1963 and 2011, shown in **FIG 10**.

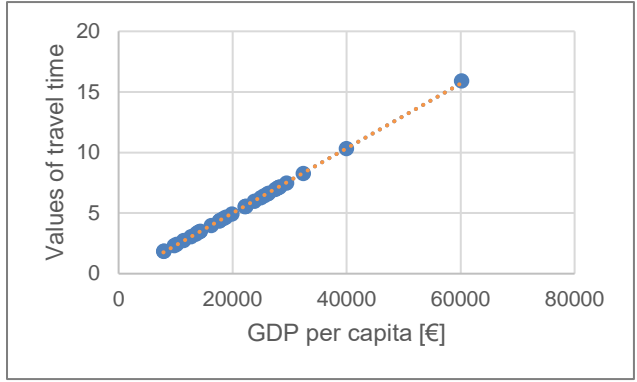

**FIG 10** Value of travel time by GDP per capita

Finally, the total air taxi demand is obtained by multiplying the air taxi modal split of each OD pair by its respective trip demand and then summing. In addition to air taxi demand, total air taxi movements and fleet size are derived. To calculate both air taxi movements and fleet size, the number of seats per aircraft, the seat load factor (SLF), and the air taxi utilization per hour are taken into account, assuming that the aircraft have four seats, the SLF is 0.5, and the utilization is 0.33 per hour.

The total number of air taxi movements per city is calculated by:

$$movements = \frac{\sum air\ taxi\ trips_{city}}{(seats\ per\ aircraft) * SLF} \quad (9)$$

The fleet size per city is calculated by:

$$fleet\ size = \frac{\sum air\ taxi\ flight\ time\ p.d._{city}}{(seats\ per\ aircraft) * SLF * utiliaztion\ p.h.} \quad (10)$$

In conclusion, all city-specific results are summed up to estimate the global demand for UAM, the global number of air taxi movements and the global fleet size.

## 4. RESULTS

The methodology described above was applied to 990 cities worldwide with populations greater than 500,000 inhabitants. In the first step, sensitivity analyses were performed to better understand the dependencies and effects of UAM specific model parameter. In a second step, different market scenarios on global UAM demand, aircraft movements and fleet size are outlined.

### 4.1. Sensitivity analyses

In the sensitivity analysis, the effects on two crucial factors, the air taxi ticket price per km and the vertiport density per sq. km are evaluated [39]. The price per air taxi km is directly related to the customers' willingness to pay and, together with travel time, is a key factor influencing the choice of transportation mode. The density of vertiports affects the time needed for access and egress and therefore has a significant impact on the total travel time.

**FIG 11** shows the global demand for UAM as a function of air taxi ticket price at constant vertiport density. At a ticket price of 2.50 € per km, demand is highest for each given vertiport density. An increase in price leads to a decrease in demand. In this case, demand decreases very steeply at first and then more gradually. This curve progression is similar for all vertiport densities. While at the lower vertiport density the demand is practically zero at an air taxi ticket price of about 3.50 € per km, at the higher vertiport density there is still demand up to a price of 4.50 € per km. This is due to the fact that at higher vertiport densities the times for pre-carriage and on-carriage are lower, making the air taxi attractive to a larger share of traffic demand even at higher prices.

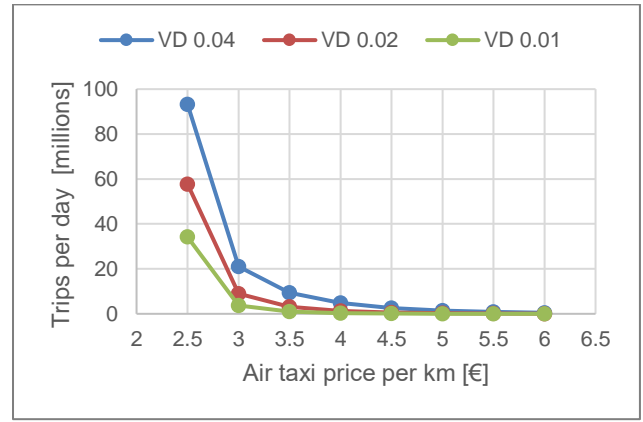

**FIG 11** Global daily UAM demand versus air taxi ticket price per km [€]

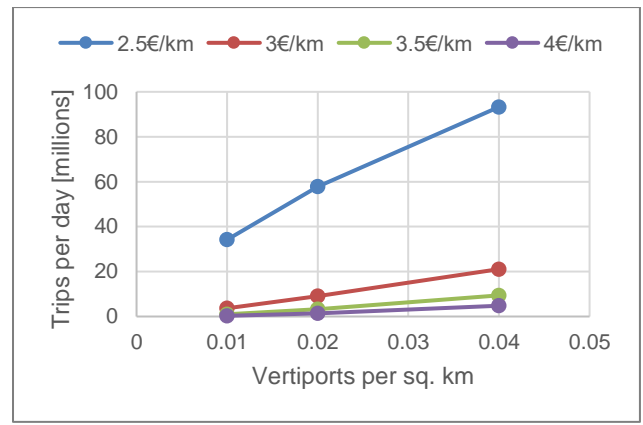

**FIG 12** Global daily UAM demand versus vertiport density

**FIG 12** shows the global UAM demand as a function of vertiport density at constant air taxi ticket prices. At low vertiport densities, demand is small and almost disappears at air taxi prices of 3.00 € per km and above. Only at an air taxi ticket price of 2.50 € per km there is a notable demand, even with a vertiport density of 0.01. This is due to the fact that the low price compensates for the relatively long pre- and post-trip times that apply to much of the demand at low vertiport densities. Thus, at constant price, demand increases when vertiport density is increased.

**Table 1** shows the results of the sensitivity analyses for the variation of air taxi ticket prices and vertiport densities.

**Table 1** Number of daily UAM trips for different air taxi ticket prices and vertiport densities

| | | Vertiport density / sq. km | | |
|---|---|---|---|---|
| | | **0.01** | **0.02** | **0.04** |
| Air taxi ticket price [€] / km | 2.50 | 34,178,038 | 57,736,491 | 93,253,541 |
| | 3.00 | 3,669,134 | 9,020,090 | 21,070,562 |
| | 3.50 | 988,826 | 3,193,060 | 9,392,581 |
| | 4.00 | 318,304 | 1,327,331 | 4,791,927 |
| | 4.50 | 108,973 | 587,926 | 2,585,372 |
| | 5.00 | 38,912 | 270,569 | 1,441,318 |
| | 5.50 | 14,342 | 127,540 | 820,061 |
| | 6.00 | 5,415 | 61,105 | 473,132 |

### 4.2. Market Development

The city-centric forecasting approach is used to outline different possible development paths of UAM until the year 2050. Four market development scenarios (S1–S4) are considered with different assumptions regarding the development of vertiport density and air taxi ticket price over time. External market conditions such as population growth and wealth development are identical throughout the scenarios.

The air taxi ticket price affects the affordability and the vertiport density determines the accessibility of UAM services, both important aspects of user acceptance. By defining a high and low vertiport density evolution and an optimistic and conservative price evolution, four market scenarios are elaborated (Table 2).

**Table 2** Vertiport density and air taxi ticket prices for the four scenarios

| | Vertiport Density | Air Taxi Prices |
|---|---|---|
| **Scenario 1** | High | Optimistic |
| **Scenario 2** | Low | Conservative |
| **Scenario 3** | High | Conservative |
| **Scenario 4** | Low | Optimistic |

The assumptions regarding air taxi ticket prices per km are based on studies by Pertz et al. (2023) [40]. Using a cost and revenue model for inner-city air taxi services, they found that under favorable conditions, an air taxi fare of 4.10 €/km is required to operate profitably. Under less favorable conditions, an air taxi fare of 5.70 €/km is required to ensure sound profitability. These values are used as a baseline for the year 2030. For further market development it is assumed that these prices decrease linearly by 1/3 until 2050, shown in **FIG 13**.

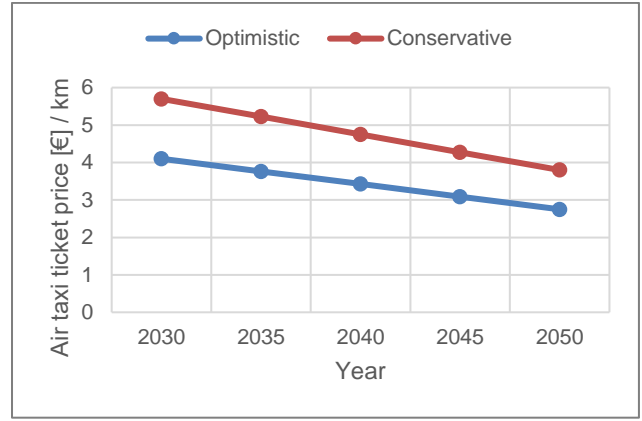

**FIG 13** Development of air taxi ticket prices over time

As the vertiport density of a city is linked to the reference vertiport density (section 3), development paths are assumed for the reference vertiport density (**FIG 14**). For the scenarios with high vertiport density, it is assumed that the reference density of 0.002 vertiports per sq. km in 2030 will increase to 0.02 vertiports per sq. km in 2050. According to Mayakonda, M., et al. (2020) [14], this corresponds to an average access and egress distance of 9 and 3 km, respectively. For the scenarios with low vertiport density, the reference vertiport density increases from 0.001 vertiports per sq. km to 0.01 vertiports per sq. km in the same period. This is equivalent to an average access and egress distance of 12 and 5 km, respectively. Thus, the potential development of vertiport density is significantly lower in the second development path.

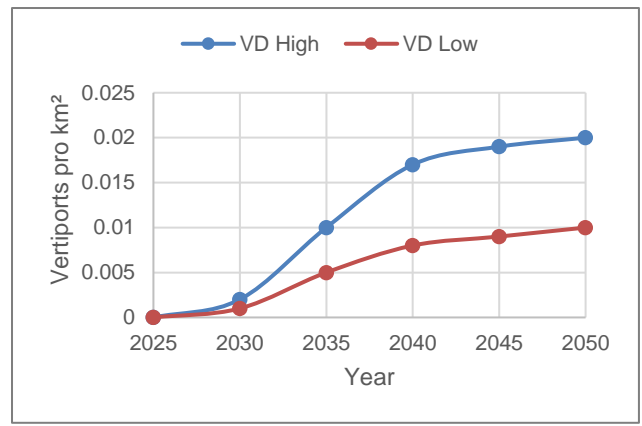

**FIG 14** Development of vertiport density over time

The evolution of the daily UAM demand, the number of movements and the corresponding fleet size for the scenarios are shown in **FIG 15**, **FIG 16** and **FIG 17**. They are similar for all four scenarios, but at different levels. Initially, the market grows very slowly in all scenarios, so that there are hardly any significant differences in the results up to 2035. From 2040 onwards, market growth increases, where scenario 1 stands out slightly from the other scenarios. From 2045 onwards, the divergence between the scenarios becomes more pronounced. Market growth is stronger in scenario 1 and 4, which are characterized by optimistic air taxi ticket prices, whereas markets develop only moderately in scenarios 2 and 3 with conservative air taxi ticket prices.

In 2050, the daily UAM demand is about 19 million passengers in Scenario 1. This demand is accompanied by about 9.7 million aircraft movements and a fleet size of about 1 million vehicles. In Scenario 4 daily UAM demand is about 9 million passengers in 2050, accompanied by about 4.5 million aircraft movements and a fleet size of about half a million vehicles.

Scenarios 2 and 3, on the other hand, market size is comparatively low by 2050. For Scenario 2 the daily UAM demand is about 400,000 passengers, with about 200,000 aircraft movements, and a fleet size of about 10,000 vehicles. By 2050, for Scenario 3 the daily UAM demand is of about 1.9 million passengers, with about 900,000 aircraft movements and a fleet size of about 50,000 vehicles.

Furthermore, the results show that there is potential for UAM in only a subset of the 990 cities under consideration, where the demand is high enough and a sufficiently large vertiport network can be set up. In 2050, the number of cities where UAM services are conceivable ranges from 135 in scenario 2 to 222 in scenario 1. Among these cities are international metropole regions such as London, Tokyo, or New York but also major German regions like the Rhine-Ruhr region, Berlin, Munich or Hamburg.

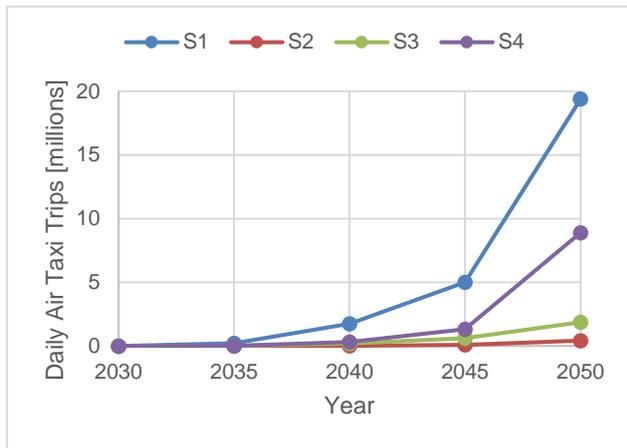

**FIG 15**  Global UAM transport demand for the various scenarios

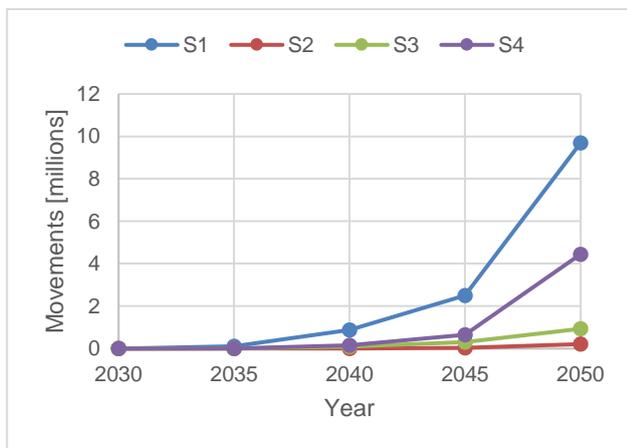

**FIG 16**  Global UAM movements for the various scenarios

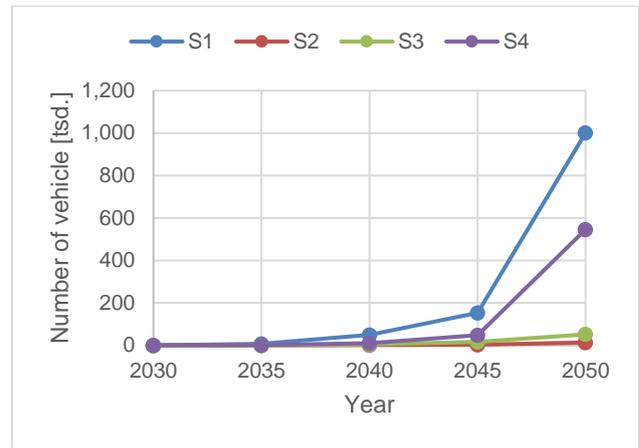

**FIG 17**  Global UAM fleet size for the various scenarios

It should be noted that the results are highly dependent on the assumptions regarding the development paths for vertiport density and air taxi ticket prices.

## 5. CONCLUSION

This paper proposes a forecasting methodology that provides initial estimates of potential global UAM demand for intra-city air taxi services. The concept is based on a city-centric approach that uses a limited number of parameters to estimate the total transportation demand for each city. A simplified multinomial logit model is used to determine the probability that travelers will choose air taxi for their individual trips within a city, using travel time and travel costs of each mode as input parameters. Based on the resulting UAM demand, cities with potential for UAM services can be identified. By summing all city-specific results, an estimate of the global demand for UAM is obtained. By varying the main characteristics of the UAM transportation system, sensitivity studies can be conducted as well as market development scenarios can be analyzed.

Sensitivity analyses were conducted to investigate the impact of vertiport density in the range of 0.01 Vertiports per sq. km and 0.04 Vertiports per sq. km as well as ticket prices between 2.50 € and 6.00 € per km on UAM demand. As expected, UAM demand is highest when air taxi ticket prices are low. However, it is remarkable how strongly demand declines as air taxi ticket price increases. UAM demand drops on a very low level when the air taxi ticket price is above about 4.00 € per km. On the other hand, demand for UAM rises with an increase in vertiport density. This implies that in order to boost the demand for UAM, either the prices should be lowered or the vertiport density should be increased. However, more vertiports usually mean higher costs and are also problematic in terms of non-user acceptance and land-use. More vertiports only make sense if they generate significantly more UAM demand, so that the costs for the additional vertiports are exceeded by the additional revenues, which needs to be further investigated in the context of a holistic cost and revenue analysis.

Considering different development paths for air taxi ticket prices and vertiport densities, four potential market development scenarios were outlined. The results show that a significant increase in UAM demand is not expected by 2040, regardless of the level of air taxi ticket prices. If air taxi ticket prices are low, demand may increase significantly by 2050, creating a mass market. However, if prices remain high, UAM demand in 2050 is likely to remain at the low level of a niche market. The results indicate that a low air taxi ticket price is more important than a high vertiport density for high demand. In the best-case scenario, a low air taxi ticket price and a high density of vertiports could result in a market potential for UAM of 19 million daily trips in over 200 cities worldwide by 2050, with a focus on North America, Europe, and Eastern Asia.

In addition, it can be concluded that the market scenarios outlined could be problematic for market introduction and require "staying power" on the part of manufacturers and operators, as the market development is characterized by low market growth in the initial phase and strong market growth thereafter. It is important to note, however, that the results shown depend on assumptions about the development paths of air taxi ticket prices and vertiport density. Lower air taxi ticket prices at the beginning of market introduction and a rapid decline in prices are conducive to market development.

In summary, the study highlights the critical role of low ticket prices and the importance of high vertiport density for fast access and egress to the UAM system to increase UAM demand. Comparing the results of this study with the findings of Pertz et al. (2023) [40], there is currently a dissonance between the air taxi ticket prices of at least 4.00 Euro per km required for profitable operations and those needed to generate high UAM demand.

This underscores the need to carefully optimize system components to minimize costs and maximize the quality of UAM services. Such an approach would contribute to the economic viability and successful deployment of UAM systems.

In conclusion, it can be stated that as long as UAM is still in the development stage, there are many uncertainties in forecasting global UAM demand. In addition, trying to make a global forecast with limited resources involves a high degree of abstraction. The forecasting approach leaves much room for further research and improvement of the proposed methodology. This includes a critical evaluation of the simplification of real urban structures to circular cities. Furthermore, the number of alternative transportation modes should be extended to distinguish between public and private transport. In this context, the parameters for the mode choice model should also be reviewed and improved. In this study, the same values were often assumed for external factors such as the number of trips per person, the distribution of trip distances, or travel speeds for different cities. Further adjustment of the data to specific city characteristics should be considered. Last but not least, the assumptions for the development of the UAM system components should be improved and integrated into the method.

The flexible design of the forecasting methodology permits the specification of all parameters, with new findings taken into account, in order to enhance long-term demand estimation for UAM and analysis of market potential across global urban areas.

## COMPETING INTERESTS

Co-Author Henry Pak is also guest editor for the special issue on the HorizonUAM project but has not been involved in the review of this manuscript.